\documentclass[english,aps,preprint]{revtex4}
\usepackage[T1]{fontenc}
\usepackage[latin9]{inputenc}
\usepackage{float}
\usepackage{bm}
\usepackage{amsmath}
\usepackage{amssymb}
\usepackage{graphicx}
\usepackage{esint}
\usepackage {ushort}

\makeatletter

\@ifundefined{textcolor}{}
{%
 \definecolor{BLACK}{gray}{0}
 \definecolor{WHITE}{gray}{1}
 \definecolor{RED}{rgb}{1,0,0}
 \definecolor{GREEN}{rgb}{0,1,0}
 \definecolor{BLUE}{rgb}{0,0,1}
 \definecolor{CYAN}{cmyk}{1,0,0,0}
 \definecolor{MAGENTA}{cmyk}{0,1,0,0}
 \definecolor{YELLOW}{cmyk}{0,0,1,0}
}

\makeatother

\usepackage{babel}
\begin{document}

\preprint{This line only printed with preprint option}

\vspace*{1cm}

\title{Ground-state wavefunction of macroscopic electron systems}

\author{P. Fulde}
\email{fulde@pks.mpg.de}
\affiliation{Max-Planck-Institut f\"ur Physik komplexer Systeme, N\"othnitzer Stra\ss e 38, 01187 Dresden, Germany}
\date{\today}

\begin{abstract}
Wavefunctions for large electron numbers $N$ are plagued by the Exponential Wall Problem (EWP), i.e., an exponential increase in the dimensions of Hilbert space with $N$. Therefore they loose their meaning for macroscopic systems, a point stressed in particular by W. Kohn. The EWP has to be resolved in order to be able to perform electronic structure calculations, e.g., for solids. The origin of the EWP is the multiplicative property of wavefunctions when independent subsystems are considered. Therefore it can only be avoided when wavefunctions are formulated so that they are additive instead, in particular when matrix elements involving them are calculated. We describe how this is done for the ground state of a macroscopic electron system. Going over from a multiplicative to an additive quantity requires taking a logarithm. Here it implies going over from Hilbert space to the operator- or Liouville space with a metric based on cumulants. The operators which define the ground-state wavefunction generate fluctuations from a mean-field state. The latter does not suffer from an EWP and therefore may serve as a vacuum state. The fluctuations have to be {\it connected} like the ones caused by pair interactions in a classical gas when the free energy is calculated (Meyer's cluster expansion). This fixes the metric in Liouville space. The scheme presented here provides a solid basis for electronic structure calculations for the ground state of solids. In fact, its applicability has already been proven. We discuss also matrix product states, which have been applied to one-dimensional systems with results of high precision. Although these states are formulated in Hilbert space they are processed by using operators in Liouville space. We show that they fit into the general formalism described above.
\end{abstract}
\maketitle

\section{Introduction}
\label{Introduction}

Electronic structure calculations for large molecules and macroscopic systems like solids remain one of the most active and challenging fields in quantum chemistry and condensed matter physics, respectively. They are pivotal for modern chemistry and of great importance for material sciences, which form the basis for many technical applications.

The first calculations of this kind started shortly after Heisenberg \cite{Heisenberg1925} and Schr\"odinger \cite{Schroedinger1926} formulated the rules for treating quantum mechanical systems. They were concerned with a deeper understanding of chemical binding. Naturally, the first system treated was the simplest one, i.e., the H$_2$ molecule. It revealed already a basic  problem, namely the proper treatment of the mutual electron repulsion. Depending on the relative size of the repulsion energy as compared with the kinetic energy of the electrons we speak of weakly or strongly correlated electrons. The early work of Heitler and London \cite{Heitler1927}, Hund \cite{Hund1928}, Mulliken \cite{Mulliken1928} and Hartree \cite{Hartree1928} stands here for many others. With increasing time the molecules which could be treated computationally increased continuously. Recently calculations of various electronic properties for molecules were reported consisting of several hundred atoms \cite{Liakos15,Werner12}.

In parallel to electronic structure calculations on molecules also those for periodic solids, i.e., macroscopic systems were performed, thereby applying more approximate techniques. For reviews see, e.g., \cite{Paulus11,Martin04,Anisinov10,Evarestov06}. Here a big obstacle is the fact that wavefunctions in Hilbert space are no longer meaningful for large, in particular macroscopic electron systems. This has been known for long time  \cite{Landau1965} and recently reemphasized by W. Kohn \cite{Kohn1999}. There are different reasons for such a statement. One is that the dimensions of Hilbert space for the description of, e.g., the ground state of a system of $N$ interacting electrons increase exponentially with $N$. This was termed by Kohn the Exponential Wall Problem (EWP). A second reason is that strictly speaking a macroscopic system can never be in a stationary state due to the interactions of the system with it surroundings. This holds true even if this interaction is extremely small like in nearly isolated systems \cite{Landau1965}.

Concerning the EWP, Kohn's arguments were summarized in the following statement \cite{Kohn1999}: for a system with $N > N_0$ electrons, where $N_0 \simeq 10^3$ a wavefunction $\psi ({\bf r}_1 \sigma_1, \dots , {\bf r}_N \sigma_N)$ is no longer a legitimate scientific concept! Here ${\bf r}_i, \sigma_i$ are the position and spin of the $i$-th electron. In order to be meaningful two conditions have to be fulfilled by a wavefunction: it must be possible to approximate it to a reasonable degree of accuracy and one has to be able to document it. Both conditions cannot be met when $N > N_0$. In this case the dimensions of Hilbert space and hence the number of parameters which have to be fixed in order to describe the wavefunction become so large that the overlap of any approximation $| \psi_{\rm app} \rangle$ to the exact ground-state wavefunction $| \psi_0 \rangle$ is zero for all practical purposes. The latter is $|\langle \psi_{\rm app} | \psi_0 \rangle | \sim (1 - \epsilon)^N$ where $\epsilon$ is a typical error in a parameter attached to a given dimension. A similar argument holds for the documentation of an exponential number of parameters, which is at least of order $2^N$.

The second argument, namely the absence of stationary states in a macroscopic electron system is based on the fact that it is never possible to decouple a system completely from its surrounding. Therefore the energy of the system is broadened by an amount of order of this interaction energy. Even when we speak of a closed system it is only quasi-closed, since in the real world there always remain some residual interactions with the surrounding. Their energy is enormous when compared with the energy level splitting in the macroscopic system. The number of levels in a given energy interval grows exponentially with particle number $N$. Therefore, the energy uncertainty due to the interactions with its the surrounding is always bigger than the energy level splitting and therefore no stationary wavefunction is strictly speaking possible. Also, it would require on astronomical time to bring a macroscopic system into a stationary state given the smallness of the energy splittings \cite{Landau1965}.

In order to resolve the problem of a proper description of the ground state of a macroscopic electron system we first neglect completely the interactions of the system with its surrounding and concentrate on the EWP. Subsequently we discuss the inclusion of the neglected interactions. When we consider the system as completely isolated, we can define stationary states in Hilbert space. Thus, the EWP remains and must be dealt with.

The EWP has its origin in the multiplicative property of wavefunctions. When $| \psi_A \rangle$ and $| \psi_B \rangle$ are the wavefunctions of two separate systems $A$ and $B$, then the wavefunction of the total system is $| \psi_{A/B} \rangle = | \psi_A \rangle \otimes | \psi_B \rangle$. The relevance of this feature for the EWP is seen by considering a system consisting of $N_A$ nearly noninteracting atoms with $n_A$ electrons each. Assume, that the correlations among the electrons on one atomic site can be described with sufficient accuracy by a superposition of $M$ electronic configurations. Then the total number of configurations required for the description of, e.g., the ground state of the total system is $M^{N_A}$ and, as expected, exponentially exploding. Yet, the information we obtain from this wavefunction is all contained in the one for a single atom. Information about extremely small interactions of electrons on different atoms is irrelevant for reasons pointed out above and need not be considered. Therefore, in order to avoid the EWP we have to find a representation of the wavefunction in which all the redundant configurations do not appear. They have to be eliminated from the beginning. Only those configurations should appear in the description of the wavefunction which provide new information about the system. This requires giving up the multiplicative character of a wavefunction and finding instead a description of wavefunctions which is {\it additive} in particular when matrix elements are calculated.

Note that the EWP does not exist for a system of noninteracting electrons or electrons for which the interactions are treated in a molecular-field approximation. In these cases the ground state of the system consists of a single configuration, e.g., a single Slater determinant or a N\'eel state. Note that an approximation to this configuration will also have a strongly reduced overlap with the exact eigenstate of $H_0$. Yet, it is {\it not} exponentially small and can be corrected by single-particle excitations. The EWP does also not occur in density-functional theory (DFT) \cite{Hohenberg1964,Kohn1965}, where all electronic variables are traced out except for those needed, e.g., for the description of the density $n({\bf r})$. This approach is a molecular-field type theory too.

The above suggests to split the Hamiltonian $H$ for a macroscopic system like a solid into two parts $H = H_0 + H_1$ with a known ground state of $H_0$, i.e., when $H_0$ is the Hamiltonian in SCF approximation
\begin{equation}
H_0 | \Phi_0 \rangle = E_{\rm SCF} | \Phi_0 \rangle~~.
\label{eq:01}
\end{equation}
The residual part $H_1$ describes the fluctuations with respect to $| \Phi_0 \rangle$. If $| \Phi_0 \rangle$ is identified with the SCF- or Hartree-Fock ground state of the electron system, then $H_1$ generates one-, two-, three-, four- etc. particle excitations out of $| \Phi_0 \rangle$. When we call $| \Phi_0 \rangle$ the {\it vacuum} state of the system, then $H_1$ generates {\it vacuum fluctuations}. In order to describe the ground-state wavefunction of a macroscopic system we have to restrict ourselves to describing those vacuum fluctuations which contain new information. This is generally not possible in Hilbert space where the wavefunction is multiplicative and contains redundant information. But it can be realized in Liouville- or operator space.

Vacuum fluctuations are described by opertors and therefore we have to consider the operator- or Liouville space.  However, the restriction to fluctuations (or operators) which contain new information, requires a special metric in Liouville space, namely one based on cumulants. We denote in the following by $| \Omega )$ the point in Liouville space which specifies the ground state of the electronic system. The rounded ket indicates that the metric in Liouville space is a special one. In order to specify $| \Omega )$ we have also to reexpress the corresponding Schr\"odinger equation in Liouville space. As we will see this poses no problem.

At this stage it should be pointed out that for extended one-dimensional systems it has been possible to determine ground-state properties with very high, i.e., machine accuracy by working seemingly in Hilbert space. Under quite general conditions \cite{Schuch08} the ground state wavefunction of a one-dimensional system can be written in form of a matrix-product state \cite{Orus14,Pollmann2016,Rommer95,Vidal07,Orus08}. High precision results for various physical properties can be obtained when for a system with a MPS an area law holds \cite{Eisert10}. Then the ground-state wavefunction does not face the EWP and we explain why this is so. Yet, calculations with matrix product states are until now feasible only for low dimensional systems, i.e., chains and in some cases two-dimensional structures. The treatment of matrix-product states is intimately related to the density-matrix renormalization group (DMRG) \cite{White1992,Peschel1999,Schollwoeck05,Fannes89,Kluemper93}. By providing a connection between MPS and the Liouville space approach we hope to stimulate discussions about possible extensions of MPS to higher dimensions.

This thematic review is structured as follows. In order to familiarize the reader with the use of cumulants we start with a brief reminder, how the free energy of a classical imperfect gas is calculated. This is followed by a summary of the most important properties of cumulants. They demonstrate their usefulness. In a next step we discuss briefly the inclusion of the coupling of a macroscopic electron system to its surrounding. Next the form of $| \Omega )$ and of the Schr\"odinger equation for $| \Omega )$ are discussed. This is followed by an incremental decomposition of $| \Omega )$, which is an important feature used in numerical applications. Finally, a list of applications of the above formalism is given. A comparison of the Liouville-space approach with the one for extended chains based on MPS completes this overview.

\section{A brief reminder: the imperfect classical gas}
\label{ImpClassGas}

It was Kubo \cite{Kubo1962} who pointed out many years ago the important role which cumulants are playing in classical and quantum statistics. They are required when multiplicative functions (e.g., the partition function, density matrices, wavefunctions etc.) are set in relation with additive  functions (e.g., free energy, densities, momenta, etc.). We demonstrate this here by choosing the partition function and the free energy of a classical gas \cite{DiCastro15}. In the following Sections we want to apply cumulants for the replacement of the wavefunction in Hilbert space, which is multiplicative by one in Liouville space, which is additive.

We denote with $Z$ the partition function of a gas which factorizes as $Z = Z_{\rm id} \cdot Z_U$. Here $Z_{\rm id}$ is the partition function of an ideal gas and $Z_U$ the modification due to the mutual interactions of the gas particles. The latter depend on the potential energy $U = \sum\limits^N_{i>j} \phi_{ij}$ of the $N$ particles with pair interactions $\phi_{ij}$. The corresponding free energy is $F(T) = F_{\rm id} + F_U$. We define
\begin{equation}
f_{ij} = \exp \left( - \beta \phi_{ij} \right) - 1~~~;~~~\beta = \left( k_B T \right)^{-1} 
\label{eq:02}
\end{equation}
and write
\begin{equation}
e^{- \beta U} = \prod\limits_{i>j} e^{- \beta \phi_{ij}} = \prod\limits_{i>j} \left( 1 + f_{ij} \right)~~.
\label{eq:03}
\end{equation}
Therefore, the interaction part $Z_U$ of the partition function is
\begin{eqnarray}
Z_U & = & \langle e^{- \beta U} \rangle \nonumber \\
& = & \langle \prod\limits_{i>j} ( 1 + f_{ij} ) \rangle
\label{eq:04}
\end{eqnarray}
where $\langle \dots \rangle$ is the average over all configurations of the gas. Consequently
\begin{equation}
F_U = - k_B T \ln \langle \prod\limits_{i>j} ( 1 + f_{ij} ) \rangle~~.
\label{eq:05}
\end{equation}
Cumulants avoid working with the logarithm of a configurational average. As discussed in the next Section, they eliminate all statistically independent, i.e., factorizable contributions to the configuration average. One definition often used is 
\begin{equation}
\ln \langle e^{\lambda A} \rangle = \langle e^{\lambda A} - 1 \rangle^c
\label{eq:06}
\end{equation}
where $A$ is an arbitrary operator or function and $c$ indicates taking the cumulant. It ensures that both sides are identical when they are expanded in powers of $\lambda$. A more general definition is given in the next Section where also their most important properties are pointed out. Here we use Eq. (\ref{eq:06}) in order to rewrite Eq. (\ref{eq:05}) in the form
\begin{eqnarray}
F_U & = & - k_B T \left< \prod_{i > j} (1 + f_{ij}) - 1 \right>^c \nonumber \\
& = & -k_B T \sum_{i>j} \left[ \langle f_{ij} \rangle + \sum_{{k<l}\atop{k,l \neq ij}} \left( \langle f_{ij} f_{kl} \rangle - \langle f_{ij} \rangle \langle f_{kl} \rangle\right) + \dots \right]~~.
\label{eq:07}
\end{eqnarray}
It demonstrates that only {\it linked} pair interactions contribute to the free energy, an observation also termed Mayer's cluster expansion \cite{Mayer1940}. The close relation of these findings with the EWP in the quantum case will become visible below.

\section{Cumulants and their properties}
\label{CumulantsProp}

For a general definition of cumulants we consider first the following function depending on $M$ parameters $\lambda_1, \dots \lambda_M$,
\begin{equation}
f \left( \lambda_1, \lambda_2, \dots, \lambda_M \right) = \ln \left< \Phi_1 \left| \prod^M_{i=1} e^{\lambda_1 A_i} \right| \Phi_2 \right> ~~.
\label{eq:08}
\end{equation}
The states $| \Phi_1 \rangle$ and $| \Phi_2 \rangle$ are non-orthogonal vectors in Hilbert space, i.e., $\langle \Phi_1 | \Phi_2 \rangle \neq 0$ and the $A_i$ are arbitrary operators. This function is analytic in the vicinity of $\lambda_1 = \lambda_2 = \dots = \lambda_M = 0$. Therefore we can expand it around this point. The expansion coefficients define the cumulants $\langle \Phi_1 | A_1 A_2 \dots A_M | \Phi_2 \rangle^c$, i.e.,
\begin{equation}
\langle \Phi_1 | A_1 \dots A_M | \Phi_2 \rangle^c  = \frac{\partial}{\partial \lambda_1} \dots\dots \frac{\partial}{\partial \lambda_M} \ln \left< \Phi_1 \left| \prod^M_{i=1} e^{\lambda_i A_i} \right| \Phi_2 \right> ~~.
\label{eq:09}
\end{equation}
For example, the cumulant $\langle \Phi_1 | A_1 A_2 | \Phi_2 \rangle^c$ is
\begin{equation}
\langle \Phi_1 | A_1 A_2 | \Phi_2 \rangle^c  = \frac{\langle \Phi_1 | A_1 A_2 | \Phi_2 \rangle}{\langle \Phi_1 | \Phi_2 \rangle} - \frac{\langle \Phi_1 | A_1 | \Phi_2 \rangle}{\langle \Phi_1 | \Phi_2 \rangle} \frac{\langle \Phi_1 | A_2 | \Phi_2 \rangle}{\langle \Phi_1 | \Phi_2 \rangle}~~.
\label{eq:10}
\end{equation}
When we require that $\langle \Phi_1 | \Phi_2 \rangle = 1$ the cumulant of the product $A_1 A_2 A_3$ is written of the form
\begin{eqnarray}
\langle A_1 A_2 A_3 \rangle^c & = &  \langle  A_1 A_2 A_3 \rangle - \langle A_1 \rangle \langle A_2 A_3 \rangle\nonumber \\
&& - \langle A_2 \rangle \langle A_1 A_3 \rangle - \langle A_3 \rangle \langle A_1 A_2 \rangle + 2 \langle A_1 \rangle \langle A_2 \rangle \langle A_3 \rangle
\label{eq:11}
\end{eqnarray}
and so on. Here the abbreviation $\langle \Phi_1 | \dots | \Phi_2 \rangle = \langle \dots \rangle$ has been introduced. Equation (\ref{eq:06}) is reproduced by setting in Eq. (\ref{eq:09}) $A_1 =  \dots = A_M = A$ multiplying with $\lambda^M/M!$ and summing over $M$.

Cumulants have the following properties, which can be easily checked \cite{Kladko1998}:\\
Linearity:
\begin{equation*}
\langle A (\alpha B + \beta C) \rangle^c = \alpha \langle AB \rangle^c + \beta \langle AC \rangle^c
\label{eq:11x}
\end{equation*}
and independence from the norm of the vectors $| \Phi_1 \rangle$ and $| \Phi_2 \rangle$:
\begin{equation}
\langle \alpha_1 \Phi_1 | AB | \alpha_2 \Phi_2 \rangle^c = \langle \Phi_1 | AB | \Phi_2 \rangle^c~~.
\label{eq:12}
\end{equation}
Products of statistically independent operators have the property that $\langle AB \rangle = \langle A \rangle \langle B \rangle$, implying that the cumulant $\langle AB \rangle^c = 0$. When two operators $A$ and $B$ are considered as an entity with respect to cumulants we denote them by $(AB)^*$ and it is generally
\begin{equation}
\langle A_1 (A_2 A_3)^* \rangle^c \neq \langle A_1 A_2 A_3 \rangle^c~~.
\label{eq:13}
\end{equation}
The cumulant of the number 1 is
\begin{equation}
\langle \Phi_1 | 1 | \Phi_2 \rangle^c = \ln \langle \Phi_1 | \Phi_2 \rangle~~.
\label{eq:14}
\end{equation}
When $\langle \Phi_1 | \Phi_2 \rangle = 1$ it follows that $\langle 1 \rangle^c = 0$ while for the unit operator $1_{\rm op}$ we find
\begin{equation}
\langle 1_{\rm op} \rangle^c =  1    \quad , \quad  \langle 1_{\rm op} \cdot A \rangle^c = 0~~.
\label{eq:15}
\end{equation}
Equation (\ref{eq:14}) is obtained by formally setting the number of $\lambda$ parameters in Eq. (\ref{eq:09}) equal to zero.

It is interesting to consider the behaviour of the cumulant when we transform the vector $| \Phi_2 \rangle$ in Eq. (\ref{eq:09}) into another vector $| \psi \rangle$ in Hilbert space. For this purpose we apply a sequence of infinitesimal transformation $e^{\delta S}$ taking us on a path in Hilbert space from $| \Phi_2 \rangle$ to $| \psi \rangle$. We subdivide this path into $L$ steps. After the first step we obtain for the cumulant of any operator $A$, but now taken with respect to the vectors $\langle \Phi_1 |$ and $e^{\delta S_1} | \Phi_2 \rangle$
\begin{equation}
\langle  \Phi_1 | A e^{\delta S_1} | \Phi_2 \rangle^c =  \langle \Phi_1 | A (1 + \delta S_1) | \Phi_2 \rangle^c ~~.
\label{eq:16}
\end{equation}
After $L$ steps this results in
\begin{equation}
\langle  \Phi_1 | A | \psi \rangle^c =  \langle \Phi_1 | A \Omega | \Phi_2 \rangle^c
\label{eq:17}
\end{equation}
with
\begin{eqnarray}
\Omega & = & \lim_{L \to \infty} \prod^L_{i=1} (1 + \delta S_i) \nonumber\\
& = & (1 + S)~~.
\label{eq:18}
\end{eqnarray}
We draw attention that $\Omega$ is not unique since many different paths can be chosen in order to go over from $| \Phi_2 \rangle$ to $| \psi \rangle$. Until now $| \psi \rangle$ has been any vector unequal $| \Phi_2 \rangle$. Later we shall choose for it the ground state $| \psi_0 \rangle$ of $H$ and for $| \Phi_2 \rangle$ the ground state $| \Phi_0 \rangle$ of $H_0$. In this case the operator $\Omega$ transforms the ground state of uncorrelated electrons into the ground state of the correlated system \cite{BeFu1989,Fulde1995,Fulde16}. Note that when $| \psi \rangle$ is any eigenstate of $H$ and $| \Phi \rangle$ is any vector in Hilbert space with $\langle \Phi | \psi \rangle \neq 0$, then for any operator $A$ (not a $c$-number!) the following equation holds \cite{Fulde1995}
\begin{equation}
\langle  \Phi | AH | \psi \rangle^c =  0~~.
\label{eq:19}
\end{equation}
The matrix element factorizes and therefore the cumulant vanishes.

\section{Ground state and Schr\"odinger equation}
\label{SchoedingerEq}

We start from a completely isolated macroscopic electron system by neglecting all interactions of the system with its surrounding. Then the Hamiltonian $H = H_0 + H_1$ of a macroscopic electron system can be written down and the ground state $| \Phi_0 \rangle$ of $H_0$ consists of a single configuration, i.e., a Slater determinant (for simplicity we assume that the ground state is nondegenerated).

We define this state as the vacuum state. As discussed before, in order to describe the wavefunction in a form which is {\it additive}, all vacuum fluctuations which enter the description of the ground state must be linked, i.e., they should not factorize. We include them by the following vector in Liouville space
\begin{equation}
| \psi_0 \rangle^c = | \Omega \Phi_0 \rangle^c~~.
\label{eq:20}
\end{equation}
This notation indicates that whenever a matrix element involving $| \psi_0 \rangle^c$ is calculated the cumulant of this matrix element must be taken. We have here adopted the notation of Eqs. (\ref{eq:17},\ref{eq:18}) and identified $| \Phi_1 \rangle$ with $| \Phi_0 \rangle$ and $| \psi \rangle$ with the ground state $| \psi_0 \rangle$ of $H$. In the following we will always assume that $\langle \Phi_0 | \psi_0 \rangle \neq 0$, although this overlap becomes exponentially small with increasing electron number $N$. Equation (\ref{eq:20}) suggests to introduce the following metric in Liouville space
\begin{equation}
(A | B) = \langle \Phi_0 | A^+ B | \Phi_0 \rangle^c
\label{eq:21}
\end{equation}
where $A$ and $B$ are arbitrary operators. The ground-state energy $E_0$ is obtained by the use of Eq. (\ref{eq:12}) with $A_1 = 1$ and $A_2 = H$ as
\begin{eqnarray}
E_0 & = & \frac{\langle \Phi_0 | H | \psi_0 \rangle}{\langle \Phi_0 | \psi_0 \rangle} \nonumber\\
& = & \langle \Phi_0 | H \psi_0 \rangle^c~~.
\label{eq:22}
\end{eqnarray}
With the help of Eq. (\ref{eq:20}) this expression is rewritten in condensed form as 
\begin{equation}
E_0 = (H | \Omega)~~.
\label{eq:23}
\end{equation}
We call $| \Omega)$ the cumulant wave operator in analogy to M{\o}ller's wave operator $\tilde{\Omega}$. The latter relates $| \psi_0 \rangle$ and $| \Phi_0 \rangle$ in Hilbert space through 
\begin{equation}
| \psi_0 \rangle = \tilde{\Omega} | \Phi_0 \rangle~~.
\label{eq:24}
\end{equation}
As seen from Eq. (\ref{eq:18}) $| \Omega )$ is of the generic form $| \Omega ) = | 1 + S )$ and therefore $| S \rangle$ is called a cumulant scattering operator. It describes those vacuum fluctuations which are connected and therefore contain new information.

Thus the energy $E_0$ decomposes into $E_0 = E_{\rm SCF} + E_{\rm corr}$ with
\begin{eqnarray}
E_{\rm SCF} & = &  (H | 1)\nonumber\\
E_{\rm corr} & = &  (H | S) = (H_1 | S)~~.
\label{eq:25}
\end{eqnarray}
The accuracy of the correlation energy $E_{\rm corr}$ depends on the quality of the description of the cumulant scattering operator $| S )$. One notices that with Eq. (\ref{eq:22}) we have gone over from a wavefunction $| \psi_0 \rangle$ in Hilbert space, which is of a multiplication form to a characterization of the ground state in Liouville space, i.e., $| \Omega )$ which is additive. There is no EWP in the latter case. Any approximation to $| \Omega )$ leads just to a small change $| \delta S )$ in the  cumulant scattering operator and a corresponding change in the correlation energy $\delta E_{\rm corr} = (H | \delta S)$. Note that Eq. (\ref{eq:23}) corresponds to the Schr\"odinger equation for the ground state formulated in Liouville space. This equation is in Hilbert space of the form of Eq. (\ref{eq:22}) with $| \psi_0 \rangle = \tilde{\Omega} | \Phi_0 \rangle$, and therefore the formulation in Liouville space is the natural one for a wavefunction with additive rather than multiplicative properties. Thus the form of Eq. (\ref{eq:23}) has to be used for macroscopic systems where the EWP invalidates the concept of wavefunctions in Hilbert space. For small electronic systems both forms, i.e., the one in Hilbert or Liouville space may be used, which ever is more convenient. Next we shall derive some relations which are very useful for practical calculations of $| \Omega )$.

We start from the identity
\begin{eqnarray}
\lim_{\lambda \to \infty} e^{- \lambda H} | \Phi_0 \rangle & = &  \lim_{\lambda \to \infty} \sum_n e^{- \lambda H} | \psi_n \rangle \langle \psi_n |  \Phi_0 \rangle\nonumber\\
& = & \lim_{\lambda \to \infty} e^{- \lambda H} | \psi_0 \rangle \langle \psi_0 |  \Phi_0 \rangle
\label{eq:26}
\end{eqnarray}
where the $| \psi_n \rangle$ are a complete set of orthonormal eigenfunctions of $H$.

We rewrite the last expression as
\begin{eqnarray}
| \psi_0 \rangle & = &  \frac{1}{\langle \psi_0 |  \Phi_0 \rangle} \lim_{\lambda \to \infty} e^{- \lambda (H - E_0)} |  \Phi_0 \rangle\nonumber\\
& = & \tilde{\Omega} |  \Phi_0 \rangle
\label{eq:27}
\end{eqnarray}
in accordance with Eq. (\ref{eq:24}). 

From Eqs. (\ref{eq:12}), (\ref{eq:20}) and (\ref{eq:27}) we conclude that
\begin{equation}
| \Omega ) = \lim_{\lambda \to \infty} | e^{- \lambda H} )~~.
\label{eq:28}
\end{equation}
The right hand side remains finite in the limit $\lambda \to \infty$. For the extraction of this remaining part we apply a Laplace transform. Note that a constant term leads to a $1/z$ contribution of the Laplace transform. Therefore by multiplying it by $z$ and taking the limit $z \to 0$ we can extract the desired term from Eq. (\ref{eq:28})
\begin{eqnarray}
\lim_{z \to 0} \frac{1}{z} | \Omega ) & = &  \lim_{z \to 0} \int\limits^\infty_0 d \lambda ~e^{z \lambda} | e^{- \lambda H} )~:~~\Re \{ z \} < 0 \nonumber\\
| \Omega ) & = & \lim_{z \to 0} \left| \left. \frac{1}{z - H} z \right) \right.
\label{eq:29}
\end{eqnarray}
The last expression can be rewritten as
\begin{eqnarray}
| \Omega ) & = &  \lim_{z \to 0} \left| \left. 1 + \frac{1}{z - H} H_1 \right) \right. \quad {\rm or} \nonumber\\
& = & \lim_{z \to 0} \sum^\infty_{n=0} \left| \left. \left( \frac{1}{z - H_0} H_1 \right)^n \right) \right.~~.
\label{eq:30}
\end{eqnarray}
We have used that $| \dots\dots H_0) = 0$ since $| \Phi_0 \rangle$ is an eigenstate of $H_0$ and therefore any cumulant vanishes. When Eq. (\ref{eq:30}) is set into Eq. (\ref{eq:25}) we obtain the energy contributions of the linked fluctuations in form of a perturbation expansion. It is equal to the Goldstone diagrammatic expansion \cite{Goldstone57} which shows that only {\it linked} diagrams contribute to $E_{\rm corr}$. From the above it is obvious that an expansion of the form of Eq. (\ref{eq:30}) holds independent of the splitting of $H$ into $H_0$ and $H_1$. Note the connections to Kato's expansions \cite{Kato80}.

While Eq. (\ref{eq:30}) enables an evaluation of $| \Omega )$ in form of a perturbation expansion, one may also adopt a quite different  approach based on projections. In case that one has a clear physical picture about the most important fluctuations one may limit oneself to these and thus to a relevant subspace $\Re_0$ of the full Liouville space $\Re$ which they span \cite{Loewdin86}. An example are a strong reduction of double occupancies of certain orbitals as compared with the vacuum, i.e., $| \Phi_0 \rangle$. Let us assume that the orthonormal operators $A_\nu$ span this subspace $\Re_0$. Then an ansatz of the form of
\begin{equation}
| \Omega ) = \left| \left. 1 + \sum_\nu \eta_\nu A_\nu \right) \right.
\label{eq:31}
\end{equation}
is suggestive. The parameters $\eta_\nu$ can be determined from Eq. (\ref{eq:19}), i.e.,
\begin{equation}
(A_\nu | H \Omega) = 0~~.
\label{eq:32}
\end{equation}
When $(A_\nu | H_1 ) \neq 0$ for all $\nu$ the equations for the $\eta_\nu$ become particularly simple. From Eqs. (\ref{eq:31},\ref{eq:32}) we obtain
\begin{equation}
(A_\mu | H_1) + \sum_\nu \eta_\nu (A_\mu | H A_\nu) = 0~~.
\label{eq:33}
\end{equation}
with the solution
\begin{equation}
\eta_\nu = -\sum_\mu L^{-1}_{\nu \mu}  (A_\mu | H_1) 
\label{eq:34}
\end{equation}
and
\begin{equation}
L_{\rho \tau} =  (A_\rho | H A_\tau) ~~.
\label{eq:35}
\end{equation}
When some of the $A_\nu$ do not couple to $H_1$, that is when for some $A_\mu$ it holds that $(A_\mu | H_1) = 0$, then these operators, respective fluctuations enter only by modifying the $\eta_\nu$ via the matrix elements $L_{\rho \tau}$. With the method of projection onto $\Re_0$, or limitation to the most important fluctuations $A_\nu$ one can easily incorporate such size extensive quantum chemical methods as the Coupled Electron Pair Approximation (CEPA-O) and variations of it \cite{Kutzelnigg1975,Meyer1971,Ahlrichs1979}. At this stage on might inquire about the relation of the present approach and the Coupled Cluster (CC) method \cite{Cizek1969,Kuemmel1978,Bishop1991}. This topic has been discussed in \cite{Schork92}, see also \cite{Fulde95_12}. The wavefunction is formulated in Hilbert space and therefore suffers from the EWP. Yet, since in the CC equations only connected fluctuations enter the correlation energy, the method is size consistent and can be used to compute energies of high quality, depending on the particular system one is dealing with.

\section{Residual interactions with the surrounding}
\label{ResidualInteractions}

At this stage we want to discuss the effect on $| \Omega )$ or $| S )$ of the interaction of the macroscopic system with its surrounding.

This coupling affects the fluctuations of additive physical quantities like the energy. We subdivide the macroscopic system into macroscopic subsystems. Then we consider the deviations of an additive physical quantity $q$ of the subsystem from its average value $\bar{q}$ when the fluctuations caused by the interaction of the subsystem with its surrounding are taken into account. It is well known that for additive physical quantities these deviations are negligible \cite{Landau1965,DiCastro15}. This is seen by determining the relative fluctuations defined by  
\begin{equation}
\Re = (\overline{(\Delta q)^2})^{1/2}/ \bar{q}  \quad ; \quad    \overline{(\Delta q)^2} =  \overline{q^2} - (\bar{q})^2
\label{eq:36}
\end{equation}
where $\bar{q}$ is the average value of $q$ and $\overline{(\Delta q)^2}$ is the mean square average of the fluctuations. Since $\overline{(\Delta q)^2} \sim N$, where $N$ is the average particle number of the subsystem and also $\bar{q} \sim N$, because $q$ is additive we find that $\Re \sim 1/ \sqrt{N}$.

As shown above, $| \Omega )$ respective $| S )$ is an additive physical quantity. The cumulant scattering operator is a sum of operators $O_\nu$ multiplied by coefficients $\alpha_\nu$, i.e., $S = \sum\limits^M_\nu \alpha_\nu O_\nu$. Their number $M$ depends on the requested accuracy, e.g., of the correlation energy a topic extensively discussed below. We are interested in the fluctuations $\delta \alpha_\nu$ of the coefficients $\alpha_\nu$ caused by the coupling of the subsystem to the surrounding. In analogy to the above, $\Re \sim 1/\sqrt{M}$ where $M > N$ because the $O_\nu$ generate a correlation hole for each of the electrons. This implies that the effect of the residual coupling of a macroscopic electron system to the surrounding on the $\alpha_\nu$ and hence on $| S )$ can be safely neglected. What remains to be done is to consider a possible effect of the residual coupling on $| \Phi_0 \rangle$, i.e., the ground state of $H_0$. Remember that the latter is used to define the vacuum. An effect of the coupling on $| \Phi_0 \rangle$ would imply changes by a noticeable amount in the molecular field contained in $H_0$. This is not the case though. Changes in the molecular field would require that the single-particle excitations  contained in the set of operators $O_\nu$ have their prefactors $\alpha_\nu$ effected by the coupling to the surrounding. However, as pointed out before all changes in the $\alpha_\nu$ coefficients are completely negligible.

Having dealt with the EWP as well as with the negligible effect of the coupling of the subsystem to its surrounding, we have a robust and solid basis for electronic structure calculations for macroscopic electron systems. What is yet missing up to here are simple rules for calculating the cumulant scattering operator $| S )$. They are derived in the following Section.

\section{Decomposition of the scattering operator}
\label{Decomposition}

After having shown that the EWP does not appear if wavefunctions are formulated in Liouville space by the fluctuations of a mean-field state $| \Phi_0 \rangle$ defined as vacuum we review briefly how the above theory is applied for realistic calculations of the ground state of solids. The formulation of the wavefunction in Liouville space puts us in a position to reduce the treatment of electronic correlations to a small number of electrons. This is done as follows.

Starting point is a set of $L$ basis function $f_i({\bf r})$ centred at different lattice sites $I, J$ etc. In terms of them the field operators $\psi_\sigma ({\bf r})$ are expressed as
\begin{equation}
\psi_\sigma ({\bf r}) = \sum^L_{i=1} a_{i \sigma} f_i ({\bf r}) \sigma~~.
\label{eq:37}
\end{equation}
For the basis functions usually orthogonalized sets of Gauss-type orbitals are chosen. In this case the corresponding creation and annihilation operators $a^+_{i \sigma}$, $a_{i \sigma}$ fulfill the anticommutation relations
\begin{equation}
\left[ a^+_{i \sigma}, a_{j \sigma'} \right]_+ = \delta_{ij} \delta_{\sigma \sigma'} \quad , ~~ \left[ a^+_{i \sigma}, a^+_{j \sigma'} \right]_+ = \left[ a_{i \sigma}, a_{j \sigma'} \right]_+ = 0~~.
\label{eq:38}
\end{equation}
The Hamiltonian expressed in terms of these operators is
\begin{equation}
H = \sum_{ij \sigma} t_{i \sigma} a^+_{i \sigma} a_{j \sigma}  + \frac{1}{2} \sum_{{ijkl}\atop{\sigma \sigma'}} V_{ijkl} a^+_{i \sigma} a^+_{k \sigma'} a_{l \sigma'} a_{j \sigma}~~.
\label{eq:39}
\end{equation}
We split the Hamiltonian into $H = H_0 + H_1$ where $H_0$ is the self-consistent field (SCF) Hamiltonian $H_{\rm SCF}$ and $H_1$ is the remaining residual interaction part $H_{\rm res}$. More explicitly, the Hamiltonian $H_1$ of the residual interactions is
\begin{eqnarray}
H_{\rm res} & = &  \sum_{ijkl} \left[ \frac{1}{2} \sum_{\sigma \sigma'}  V_{ijkl} a^+_{i \sigma} a^+_{k \sigma'} a_{l \sigma'} a_{j \sigma} \right. \nonumber\\
&& \left.  - \sum_\sigma \left( V_{ijkl} - \frac{1}{2} V_{ilkj} \right) P_{kl} a^+_{i \sigma} a_{j \sigma} + \frac{1}{2} \left( V_{ijkl} - \frac{1}{2} V_{ilkj} \right) P_{ij} P_{kl} \right]
\label{eq:40}
\end{eqnarray}
where $P_{ij}$ is the density matrix
\begin{equation}
P_{ij} =  \left< \Phi_{\rm SCF} \left| a^+_{i \sigma} a_{j \sigma} \right| \Phi_{\rm SCF} \right>
\label{eq:41}
\end{equation}
and $| \Phi_{\rm SCF} \rangle$ is the ground state of $H_{\rm SCF}$. We will still use the notation $| \Phi_0 \rangle$ and $H_1$ and switch to $| \Phi_{\rm SCF} \rangle$ and $H_{\rm res}$ only when for reason of clarity this is required.

The SCF ground-state $| \Phi_0 \rangle$ which we here call {\it vacuum state} is usually written in the form of $| \Phi_0 \rangle = \prod\limits_{\mu \sigma} c^+_{\mu \sigma} | 0 \rangle$, where the $c^+_{\mu \sigma}$ create electrons in the canonical SCF or Bloch spin orbitals $\mu \sigma$. The index $\mu$ includes the momentum ${\bf k}$ and a subband index while $| 0 \rangle$ is the empty state. The vacuum fluctuations generated by $H_1$ are rather local and  generate the correlation hole of an electron. Therefore it proves advantageous to replace occupied Bloch orbitals by Wannier orbitals. The latter are obtained by a unitary transformation $U$ in the space spanned by the {\it occupied} canonical spin-orbitals
\begin{equation}
\tilde{c}^+_{\nu \sigma} = \sum^{N/2}_{\mu=1} U_{\nu \mu} c^+_{\mu \sigma}~~,
\label{eq:42}
\end{equation}
so that $| \Phi_0 \rangle = \prod\limits^N_{\nu \sigma} \tilde{c}^+_{\nu \sigma} | 0 \rangle$. The unitary transformation is chosen so that the Wannier orbitals are as localized as possible. For different localization procedures of which the one of Foster and Boys \cite{Foster60} and Edmiston and Ruedenberg \cite{Edmiston63} are the most wide spread ones we refer to the original literature. The unoccupied or virtual SCF spin orbitals are best expressed in terms of $\tilde{a}^+_{i \sigma} (I)$, $\tilde{a}_{i \sigma} (I)$ operators. They are referring to the modified basis function $\tilde{f}_i ({\bf r})$ which are the $f_i ({\bf r})$ orbitals but orthogonalized to the occupied space, i.e., to the Wannier orbitals. The index $I$ indicates the site (or bond) at which the virtual orbitals are centered. With these definitions the residual interactions can be decomposed in the form
\begin{equation}
H_1 = \sum_I H_I + \sum_{\langle IJ \rangle} H_{IJ} + \sum_{\langle IJK \rangle} H_{IJK} + \sum_{\langle IJKL \rangle} H_{IJKL}~~.
\label{eq:43}
\end{equation}
The brackets refer to pairs, triplets and quadrupoles of sites or bonds. The residual interaction part of $H$ has one or two destruction- and creation operators and the subscrips $I$, $IJ$ etc specify where these two or four operators are centered. For example, $H_I$ tells us that they are all centered at site (bond) $I$, while $IJ$ implies they are centered at sites (bond) $I$ and $J$ and so on.

Equation (\ref{eq:30}) suggests the introduction of operators
\begin{equation}
A_\alpha = \lim_{z \to 0} \frac{1}{z-H_0} H_\alpha
\label{eq:44}
\end{equation}
with $\alpha$ running over all contribution to $H_1$, i.e., $H_I$, $H_{IJ}$, $H_{IJK}$, $H_{IJKL}$. Thus from the expansion (\ref{eq:30}) we obtain
\begin{eqnarray}
| S ) & = & \lim_{z \to 0} \left. \left| \sum^\infty_{n=1} \left( \frac{1}{z - H_0} H_1 \right)^n  \right. \right) \nonumber\\
& = & \sum^\infty_{n=1} \left. \left| \left( \sum_\alpha A_\alpha \right)^n \right. \right)~~.
\label{eq:45}
\end{eqnarray}
This form is very suitable for the determination of the most important increments to $| S )$ and the correlation energy $E_{\rm corr} = (H_1 | S)$. In general correlation-energy contributions from $H_I$, i.e., from electrons on a given site $I$ will be more important than from electrons on different sites, i.e., $H_{IJ}$. Also the correlation energy contributions are expected to decrease as the sites $I$ and $J$ increase their distance. Thus the following ordering of the various terms in (\ref{eq:44}) suggests itself
\begin{eqnarray}
| S ) & = & \sum_\alpha \left. \left| \left( \sum^\infty_{n=1} A^n_\alpha \right)  \right. \right) + \sum_{\alpha \neq \beta} \left. \left| T_{\alpha \beta} \right. \right> \nonumber\\
& = & \sum_\alpha | S_\alpha ) + \sum_{\alpha \neq \beta} \left. \left| T_{\alpha \beta} \right. \right>~~.
\label{eq:46}
\end{eqnarray}
Obviously the operator $| S_\alpha )$ is the cumulant scattering operator of a Hamiltonian $H_0 + H_\alpha$. The remaining part in Eq. (\ref{eq:45}) consists of operators $| T_{\alpha \beta} \rangle$ involving more than a single $H_\alpha$. A discussion of the $| T_{\alpha \beta} \rangle$ in found in Ref. \cite{Kladko1998}.

The largest contributions come without doubts from the $| S_\alpha )$, when $\alpha$ refers to one of the  $H_I$, i.e., a single site or bond. If these are the only contributions to $| S )$, i.e., if $| S ) = \sum\limits_I | S_I )$ we speak of a {\it single-center approximation}. In this case $| S_I )$ is the cumulant scattering matrix of a Hamiltonian $H_{1c} = H_{\rm SCF} + H_I$ and the correlation energy is determined from
\begin{equation}
E_{\rm corr} = \sum_I \left( H_I | S_I  \right)~~.
\label{eq:47}
\end{equation}
Treating $H_{1c}$ is a many-body problem involving a small electron number only, i.e., those at site $I$. 

It is well known that for strongly correlated electrons a SCF ground state $| \Phi_0 \rangle$ is a poor starting point. This can be improved at this stage: By freezing all electrons in $| \Phi_0 \rangle$ except those centered at site (or bond) $I$, we can include in $| S_I )$ strong correlations of electrons on this site, if required. Strong on-site correlations can be treated, e.g., by a complete active space SCF calculation (CASSCF) of the electrons at site $I$. With $| \Omega ) = | 1 + \sum\limits_I S_I )$ they are taken into account at all sites. In Hilbert space such a generalization is not possible, as this would require to deal with an exponential number of configurations.

In an improved approximation two-site scattering matrices are included. This implies including not only $H_I$ but also $H_{IJ}$ when the cumulant scattering matrix $| S)$ is determined. This is called a {\it two-center approximation}. Since the index $\alpha$ runs now over all interaction matrix elements involving sites $I$, $J$, and $IJ$ we note that $| S_\alpha )$ contains also contributions of the form $| S_{IJ} )$.

The operator
\begin{equation}
| S_{IJ} ) = \sum^\infty_{n=1}  \left. \left| \left( A_I + A_J + A_{IJ} \right)^n \right. \right)
\label{eq:48}
\end{equation}
is then the cumulant scattering operator of the Hamiltonian $H_{tc} = H_{\rm SCF} + H_I + H_J + H_{IJ}$, i.e.,
\begin{equation}
| S ) = \sum_I | S_I ) + \sum_{\langle IJ \rangle} \left. \left| S_{IJ} - S_I - S_J \right. \right)~~.
\label{eq:49}
\end{equation}
In the two-center approximation the correlation energy is obtained from
\begin{equation}
E_{\rm corr} = \sum_I  \left( H_1 | S_I \right) + \sum_{\langle IJ \rangle} \left( H_1 | \delta S_{IJ} \right)
\label{eq:50}
\end{equation}
with $| \delta S_{IJ} ) = | S_{IJ} - S_I - S_J )$. All electrons in $| \Phi_0 \rangle$ are kept frozen except for those centered at sites (or bonds) $I$ and $J$. They are permitted to fluctuate. 

The expansion of $| S )$ can be continued so that the cumulant scattering matrix involves an increasing number of sites on which the electrons fluctuate
\begin{equation}
| S ) = \sum_I | S_I ) + \sum_{\langle IJ \rangle} | \delta S_{IJ} ) + \sum_{\langle IJK \rangle} | \delta S_{IJK} ) + \dots
\label{eq:51}
\end{equation}
with $| \delta S_{IJK} ) = | S_{IJK} - \delta S_{IJ} - \delta S_{IK} - \delta S_{JK} - S_I - S_J -S_K )$ etc.

\begin{figure}[H]
\begin{centering}
A\includegraphics[width=0.42\textwidth]{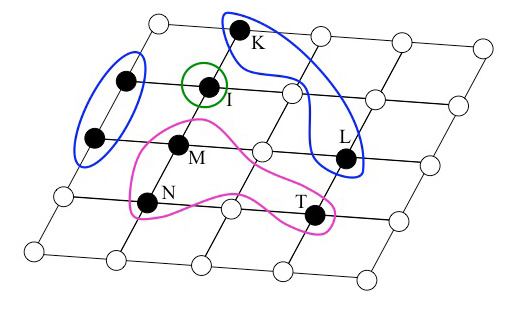}
B\includegraphics[width=0.4\textwidth]{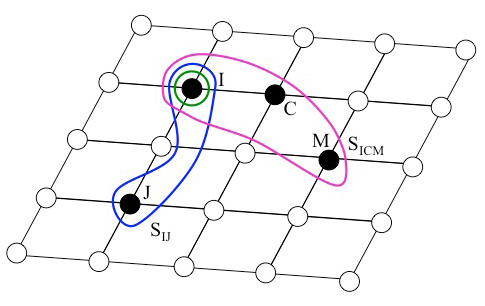}
\par\end{centering}
\caption{1a: Examples of different vacuum fluctuations $S_I$, $S_{KL}$, $S_{MNT}$ contributing to $| S )$.\\
1b: Vacuum fluctuations contributing to the correlation hole around site $I$. Different colours refer to vacuum fluctuations involving electrons on different numbers of sites.
\label{fig:01}}
\end{figure}
In Fig.\ref{fig:01} we show examples of different terms in Eq.~(\ref{eq:51}). They represent various vacuum fluctuations. They take place in Fig.~\ref{fig:01}a at different sites and in Fig.\ref{fig:01}b around site $I$. The associated correlation energy improves rapidly with increasing number of increments \cite{Stoll1992,Stoll1992a}.

The decomposition of $| S )$ and with it the computation of the correlation energy in form of increments has reduced the computations for a macroscopic system to one of a few electrons. Which of the different quantum chemical methods is the most economical one to treat these electrons depends on the special system. Often a CC or a CEPA calculation will be the method of choice. A special comment with respect to metals is in order. Here we deal with the difficulty that in a metal the occupied Wannier functions fall off only algebraically in distinction to the exponential drop in systems with an energy gap \cite{Kohn1973}. One way to improve localization here is to define from the occupied canonical or Bloch orbitals only as many localized orbitals that each of these can be doubly occupied. In the case of $Li$ metal this implies that the localized orbitals $\phi_i$, determined, e.g., by the method of Pipek and Mezey \cite{Pipek89} are set up with respect to $Li_2$ units \cite{Stoll2009}. When $\tilde{c}^+_{i \sigma}$ is the creation operator of an electron with spin $\sigma$ in orbital $\phi_i$, the SCF ground state is again of the form
\begin{equation}
| \Phi_0 \rangle = \prod^{N/2}_{i \sigma} \tilde{c}^+_{i \sigma} | 0 \rangle
\label{eq:52}
\end{equation}
where $N$ is the number of electrons in the half-filled conduction band. The different increments to $| S )$ are calculated by using the $\tilde{c}^+_{i \sigma}$ operators \cite{Stoll2009}. The more general procedure is to project the localized orbitals $\phi_i$ onto the occupied as well as onto the unoccupied, i.e., virtual SCF space. These projections $\{P_{\rm occ} \phi_i\}$ and $\{P_{\rm virt.} \phi_i\}$ are, of course, nonorthogonal and overcomplete: The overcompleteness can be remedied by finding pairs $\{\phi_i, \phi_j\}$ with largest overlap and eliminating from $\{P_{\rm occ} \phi_i\}$ and $\{P_{\rm occ} \phi_j\}$ one of the two. This is done until the overcompleteness is removed. Afterwards the remaining orbitals are pairwise orthogonalized. One notices that this leads in the case of $Li$ metal with cubic structure precisely to the procedure described above.

\section{Application}
\label{Application}

The main purpose of this communication is to present a solid basis for ground-state calculations based on wavefunctions when the electron systems we deal with are macroscopic. Yet, it is assuring to see that the theory can and has been successfully applied to solids and therefore we want to mention a number of applications which have been made. When one consults the original literature for the given examples, one will notice that the calculations described there are often using a somewhat different language. This is not surprising since the condensed form presented here of resolving the EWP problem has been developing over the years. However, the essence of the applied computational schemes in the given examples is precisely the same as described here.

Ground-state calculations have been performed for semiconductors of group IV \cite{Paulus2006}, III - V \cite{Paulus2006,Paulus96}, II - VI \cite{Albrecht92} compounds, on oxides MgO \cite{Doll95}, and CaO \cite{Doll96} to name a few. Also the rare-earth compound GaN \cite{Kalvoda98} has been treated with the $4f$ electrons kept in the core. The accuracy of the results, e.g., for the cohesive energy or the bulk modulus has been analysed in detail for some of these systems with good results \cite{Paulus11,Stoll05}.

The overall impression is that connected vacuum fluctuations are of rather small spatial extent! For example, the correlation energy due to two-body increments $| S_{IJ} )$ falls off asymptotically like van der Waals interactions do, i.e., like $r^{-6}$. They model the correlation hole around an electron. For distances larger than twice the radius of the correlations hole, electrons behave nearly as independent of each other. An analysis shows that one- and two-center correlations are usually sufficient to obtain satisfactory results for quantities like the cohesive energy, bulk modulus or bond length. This assumes that reasonably sized basis sets of Gaussian type of orbitals (GTO) are used. The influence of the size of the basis sets on the quality of the calculated physical quantities is also discussed, e.g., in Refs. \cite{Paulus11,Stoll05}. A general finding is that large energy gaps lead to spatially reduced correlations holes.

Rare-gas solids are special, since binding is not obtained on a SCF level. In this case $H_0$ is chosen so that it describes a collection of free atoms which are considered as the vacuum. The Hamiltonian $H_1$ and with it the vacuum fluctuations take care for the interactions between them \cite{Rosciszewski99}. The decomposition of $| S )$ starts therefore with the contributions $| S_{IJ} )$ where the indices refer to different atoms. They lead to binding and are at large distances of van der Waals type. By including three-body corrections of the form $| S_{IJK} )$ the accuracy of the calculated cohesive energy can be improved. We refrain from a more detailed discussion here, since the central issue of the paper is to address the more general problem of resolving the EWP.

\section{Matrix-Product States}
\label{Matrix}

As mentioned in the Introduction wavefunctions in form of matrix-product states can give highly accurate results for one-dimensional macroscopic electron systems, despite that all calculations are done in Hilbert space. This might seem puzzling since due to the EWP the concept of wavefunctions looses its meaning in Hilbert space for $N \geq 10^3$. So does this limitation not hold for macroscopic chains? The answer is: for any macroscopic interacting electron system the overlap of the exact ground-state wavefunction with any approximate form of it is exponentially small. However, for systems with an area law one need not account for all possible correlations, e.g., of spins in a spin chain. Instead one starts, e.g., from a molecular-field ground state such as a N\'eel state for a Heisenberg chain, and improves or upgrades it stepwise by means of properly chosen operators, i.e., by elements of Liouville space. When we define the initial configuration again as our vacuum, then these operators generate vacuum fluctuations and the similarity with Section \ref{SchoedingerEq} becomes obvious. But in the special case of one dimension and when the Hamiltonian contains local interactions these vacuum fluctuations can be chosen so that the stepwise upgradings are the same everywhere in the chain and they are also connected. Therefore one may remain in Hilbert space and need not introduce a cumulant metric in Liouville space. These features become most transparent when the upgrading is done with the method of Infinite Time Evolution Block Decimation (iTEBD). The method is equivalent to the DMRG which seemingly is more used in applications. Before a more detailed discussion is given we have to recall some basic features of MPS \cite{Chou12}. We start with a chain of $L \gg 1$ sites and $N$ electrons. In the simplest case of one orbital per site, each site $n$ can be in four different configurations, i.e., empty, singly occupied with a spin up (down) electron or doubly occupied. More generally, a given site $n$ can be in $d$ different configurations and $| j_n \rangle$ denotes the corresponding $d$ dimensional basis. Any wavefunction can therefore be written in the form
\begin{equation}
| \psi \rangle = \sum_{j_1, \dots j_L} C_{j_1 \dots j_L} \left. \left| j_1, j_2, \dots, j_L \right. \right>
\label{eq:53}
\end{equation}
and there are $d^L$ parameters $C_{j_1 \dots j_L}$ reduced by the requirement of a fixed electron number and total spin. Without loss of generality the matrix ${\mathcal C}$ of rank $L$ can be rewritten in form of a sum of matrix products \cite{Peschel1999,Schollwoeck05,Orus14,Pollmann2016,Rommer95,Rommer97}
\begin{equation}
C_{j_1 \dots j_L} = \sum_{\{\alpha_n\}} A [1]^{j_1}_{\alpha_1} \cdot A [2]^{j_2}_{\alpha_1 \alpha_2} \cdot A [3]^{j_3}_{\alpha_2 \alpha_3} \dots \dots A^{j_L}_{\alpha_L - 1}~~.
\label{eq:54}
\end{equation}
The sum is over all coefficients $\alpha_n$. This factorized form defines a MPS, here with open boundary conditions. The matrices ${\pmb A}[n]$ are rank-3 tensors. The upper index $j_n$ labels the $d$ configurations at site $n$, while the lower two indices $\alpha_{n-1}$, $\alpha_n$ are called bond indices and specify the bond dimensions. The step from Eqs. (\ref{eq:53}) to (\ref{eq:54}) follows from a sequence of Schmidt decompositions of the wavefunction $| \psi \rangle$ \cite{White1992,Peschel1999,Schollwoeck05,Orus14,Pollmann2016,Rommer95}. In a Schmidt decomposition the chain is cut into two parts $L$ (left) and $R$ (right). Thereby the Hilbert space $\mathcal{H}$ is divided into two parts $\mathcal{H} = \mathcal{H}_L \otimes \mathcal{H}_R$. The two parts of the chain are built from vectors $| \alpha \rangle_L$ and $| \alpha \rangle_R$ in the corresponding spaces $\mathcal{H}_L$ and $\mathcal{H}_R$, respectively. Thus $|\psi \rangle$ can be written as
\begin{equation}
|\psi \rangle = \sum_\alpha \lambda_\alpha |\alpha \rangle_L \otimes |\alpha \rangle_R~~.
\label{eq:55}
\end{equation}
The real coefficients $\lambda_\alpha$, named Schmidt coefficients obey the sum rule $\sum\limits_\alpha \lambda^2_\alpha = 1$ and are a measure of the entanglement of the two parts of the chain. In case that they are unentangled, i.e., when the electrons on the right part are uncorrelated with the ones on the left part, there remains only one Schmidt number $\lambda = 1$ in Eq. (\ref{eq:55}). By consecutive Schmidt decompositions along the chain the matrices $A^{[n] j_n}_{\alpha_{n-1}, \alpha_n}$ can be rewritten in the form \cite{Peschel1999,Schollwoeck05,Orus14,Pollmann2016,Rommer95}
\begin{equation}
{\pmb A}[n]^{j_n} = {\pmb \Gamma} [n]^{j_n}  {\pmb \Lambda} [n]
\label{eq:56}
\end{equation}
where the $\pmb{\Gamma} [n]^{j_n}$ are matrices of dimension $\alpha_{n-1} \times \alpha_n$ and the $\pmb{\Lambda} [n]$ are diagonal square matrices of dimension $\alpha_n$. The entries of there matrices are $\alpha_n$ Schmidt coefficients $\lambda_1, \dots, \lambda_{\alpha_n}$. Remember that $\alpha_n$ is the bond index of site $n$. With this replacement we obtain the canonical form of the MPS \cite{Vidal07,Vidal03}
\begin{equation}
| \psi \rangle = \sum_{j_1, \dots , j_2} \pmb{\Gamma} [1]^{j_1}  \pmb{\Lambda} [1] ~ \pmb{\Gamma} [2]^{j_2}  \pmb{\Lambda} [2] \dots \dots \pmb{\Gamma} [L]^{j_L} | j_1, j_2, \dots , j_L \rangle~~.
\label{eq:57}
\end{equation}

\begin{figure}[H]
\begin{centering}
\includegraphics[width=0.7\textwidth]{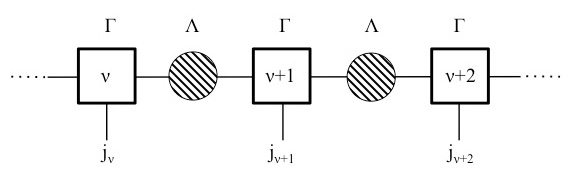}
\par\end{centering}
\caption{Graphical representation of the canonical form of  a MPS
\label{fig:02}}
\end{figure}

\noindent The conventional graphical representation of this wavefunction is shown in Fig.\ref{fig:02}. The bond order and with it the number of Schmidt coefficients increases exponentially like $d^n$ as the site number $n$ increases. This number is counted from a chain end. In order to terminate this increase, one orders the Schmidt coefficients according to their size and keeps only the $D$ largest of them. This way the bond dimension $\alpha_n$ remains constant when $d^n \gtrsim D$. Thus
\begin{equation}
\dim \alpha_n = \min \left( d^n, d^{L-n}, D \right)~~.
\label{eq:58}
\end{equation}
This cut neglects correlations between particles which are too small and eliminates the EWP. Typically $D$ is chosen so that it is reached after a few steps, e.g., $D = d^6$ Of course, when a system approaches an electronic phase transition $D$ has to increase correspondingly.

In the following we will use the canonical form for further considerations. In order to be more specific we consider a macroscopic chain with sites $I$ and a local interaction Hamiltonian for the electrons of the form $H = \sum\limits_I H_{I, I+1}$. This Hamiltonian applies to a number of spin systems and we will have in the following a Heisenberg antiferromagnet $H = J \sum\limits^{L-1}_{I=1} {\bf S}_I {\bf S}_{I+1}$ in mind. The aim is to determine the ground-state wavefunction, which is commonly written as $| \psi_0 \rangle$ in Hilbert space, although for macroscopic systems it is not a legitimate concept. Yet, what is a valid concept is to start from a N\'eel state $| \Phi_0 \rangle$, consider it as the vacuum state and to improve the description of the ground state by including vacuum fluctuations generated by the residual interactions. The N\'eel state is written in form of a canonical MPS with $\Gamma [{n={\rm even}}, \uparrow] = \Gamma [{n={\rm odd}}, \downarrow] = 1$ and $\Gamma [{n={\rm even}}, \downarrow] = \Gamma [{n={\rm odd}}, \uparrow] = 0$. Furthermore, the Schmidt coefficient is $\lambda = 1$ because it is a mean-field state as explained before. The N\'eel state is improved or upgraded by applying the infinite Time Evolving Block Decimation (iTED) according to which
\begin{eqnarray}
| \psi_0 \rangle & = & \lim_{\lambda \to \infty} \frac{e^{- \lambda H} | \Phi_0 \rangle}{\langle \Phi_0 (\lambda) | \Phi_0 (\lambda) \rangle^{1/2}} \qquad ; \quad | \Phi (\lambda) \rangle = e^{- \lambda H} | \Phi_0 \rangle \nonumber \\[3ex]
& = & \lim_{\lambda \to \infty} U(\lambda) | \Phi_0 \rangle
\label{eq:59}
\end{eqnarray}
This should be compared with the analogous formulations in Liouville space where the vacuum $| \Phi_0 \rangle$ is the same as above and where $| \psi_0 \rangle$ is replaced by $| \Omega )$. The fluctuations are determined from Eq. (\ref{eq:28}), i.e., $| \Omega ) = \lim\limits_{\lambda \to \infty} | e^{- \lambda H})$. 

\begin{figure}[H]
\begin{centering}
\includegraphics[width=0.7\textwidth]{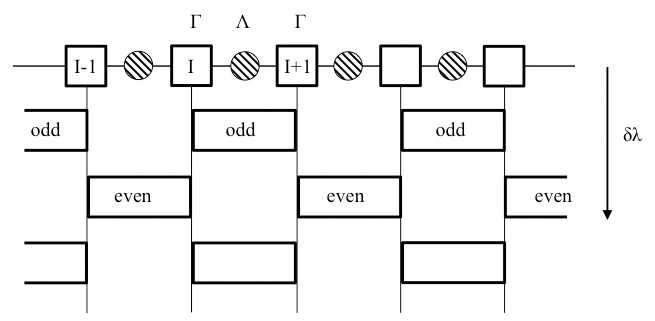}
\par\end{centering}
\caption{Starting from a mean-field state $| \Phi_0 \rangle$ simultaneous updates an every other bond (site $I$ is odd) are being made. They are done according to $U (\delta \lambda)$ with a local Hamiltonian $H = \sum\limits_I H_{I, I+1}$.
\label{fig:03}}
\end{figure}

For the upgrading with Eq. (\ref{eq:59}) we divide $H$ into $H = \sum\limits_{I={\rm even}} H_{I, I+1} + \sum\limits_{I={\rm odd}} H_{I, I+1}$. This has the advantage that the terms with $I={\rm even}$ compute with each other and so do the terms with $I={\rm odd}$. The sum in the exponent of $e^{- \lambda H}$ can be converted into a product of exponentials by means of the Suzuki-Trotter expansion \cite{Suzuki76}. It its simplest form it is written as
\begin{equation}
e^{\delta \lambda (A+B)} = e^{A \delta \lambda} e^{B \delta \lambda} + O ((\delta \lambda)^2)
\label{eq:60}
\end{equation}
when $A$ and $B$ are noncommuting operators. When applied to the present situation we obtain
\begin{equation}
e^{-\delta \lambda H} = \prod_{I={\rm even}} \exp (- \delta \lambda H_{I, I+1}) \prod_{I={\rm odd}} \exp (- \delta \lambda H_{I, I+1}) + O ((\delta \lambda)^2)~~.
\label{eq:61}
\end{equation}
The smaller $\delta \lambda = \lambda/M$ is with $M \gg 1$, the better works the decomposition. Methods of reducing the errors are found, e.g., in Ref. \cite{Hastings09}. Equation (\ref{eq:61}) can be used to first upgrading simultaneously all bonds with $I={\rm odd}$. This is possible because, as pointed out before, the different operators commute. The upgrading changes the matrices $\pmb{\Gamma} [I]$ and $\pmb{\Gamma} [I+1]$ into $\tilde{\pmb{\Gamma}} [I]$ and $\tilde{\pmb{\Gamma}} [I+1]$. Also $\pmb{\Lambda} [I]$ is changed to $\tilde{\pmb{\Lambda}} [I]$ The new matrices are obtained from a Schmidt decomposition of bond $I$. Because of the simultaneous upgrading at all $I={\rm odd}$ sites, the matrices are modified everywhere in the same way. In a next step all bonds with $I={\rm even}$ are upgraded, now with the new $\tilde{\Gamma}$ and $\tilde{\Lambda}$ matrices. Again, the upgrading is done in parallel for all $I={\rm even}$ bonds. This is shown schematically in Fig.~\ref{fig:03}.

Note that the matrix $U(\lambda)$ is not unitary and therefore the canonical form of a MPS is not conserved after an upgrading. Yet, it turns out that the deviation from the canonical form can be kept sufficiently small and therefore may be neglected \cite{Vidal07}. The upgrading is repeated until the energy calculated with the upgraded wavefunction has the required accuracy. An exponential increase of the bond order with increasing number of imaginary time steps is prevented by using Eq. (\ref{eq:58}). For more details of the upgrading procedure we refer to the original literature \cite{Vidal07}. The point we want to make here is that the upgrades (or alternatively the fluctuations) of the N\'eel state $| \Phi_0 \rangle$ are {\it additive}. They are also connected, i.e., the operators involved in an upgrading with an imaginary time step $\delta \lambda$ connect with operators involved in the previous one. Cumulants need not be introduced here. When we start from an unentangled mean-field state, the sequence of upgrading steps never generates unentangled parts of a chain. Otherwise cumulants would have to exclude them. 

This explains why calculations with MPSs in Hilbert space are successful despite the EWP. By starting from a mean-field state which is well defined in Hilbert space, the operators generating the fluctuations in MPS's are additive and connected like in a Liouville space with cumulant metric. Note the similarity to the treatment of $| \Omega )$ in Liouville space with cumulant metric. Yet, the routes taken in the two approaches are different. In the MPS scheme the ground-state is approached through the operator $e^{-\lambda H}$ (with $\lambda$ sufficiently large) by a sequence of small steps $\delta \lambda$. On the other hand, in $| \Omega )$ the limit $\lambda \to \infty$ is taken directly by  a Laplace transform (see Eq. (\ref{eq:29})) and the ground state is approached in form of an expansion in powers of $(z - H_0)^{-1} H_1$. A more detailed comparison of ground-state energy calculations for a Heisenberg chain by applying the two methods will be the subject of a separate paper \cite{Javanmard18}. An application of cumulants to a Heisenberg Hamiltonian on a square lattice is found in \cite{Becker89}.

\section{Summary and Conclusions}
\label{Summary}

The aim of this review has been to address and resolve the exponential wall problem which one is facing in Hilbert space for wavefunctions of macroscopic systems of interacting electrons. The exponential increase of the dimensions in Hilbert space with electron number renders the concept of wavefunctions obsolete in this particular space. For all practical purposes any approximate wavefunction has zero overlap with the exact one. This problem must be resolved in order to perform wavefunction based electronic structure calculations for solids. As was demonstrated it is the multiplicative property of a wavefunction with respect to independent subsystems which is causing the EWP. Therefore it is avoided when we formulate the wavefunctions so that they are additive instead of multiplicative. This is possible by choosing a wavefunction in mean-field approximation as a vacuum state and by using the operators which generate vacuum fluctuations  through $H_1$ for the definition of the ground-state wavefunction $| \Omega )$. These fluctuations define a vector in operator- or Liouville space. However, for the wavefunction to be additive a cumulant metric in Liouville space is required. The logarithm of a multiplicative function changes it into an additive one and cumulants avoid dealing with the logarithm. As a good example serves a classical interacting gas where the logarithm of the multiplicative partition function changes it into an additive function proportional to the free energy and where working with the logarithm is avoided by a cumulant expansion (Mayer's cluster expansion) of the pair interactions. Thus for a macroscopic electron system we may start from a self-consistent field, e.g., Hartree-Fock ground state and use the cumulant scattering operator $| S )$ to define the ground-state wavefunction in Liouville space through $| \Omega ) = | 1 + S )$. The round ket refers to the cumulant metric. As explained in the text $| \Omega )$ does not suffer from the EWP and provides a solid basis for electronic structure calculations for the ground state of solids. Expectation values of operators $A$ in the ground state of the system are obtained from $\langle A \rangle_{\rm exp} = (\Omega | A \Omega)$. With the help of an incremental decomposition the different contributions to the cumulant scattering operator and the correlation energy $E_{\rm corr} = (H|S)$ can be determined. Examples for the application of wavefunction based electronic structure calculations for solids do exist and were pointed out.

Special attention has been devoted to Matrix Product States. They apply mainly to one-dimension and are formulated in Hilbert space. Highly accurate results for spin chains have been obtained with them. We dealt with the question why the EWP does not plague calculations with MPSs. With the use of iTED the following has been shown: starting from a mean-field ground-state wavefunction in form of a MPS, the improvements or upgradings of this wavefunction are done with operators which are additive like different contributions to $| S )$ and connected. They define a point in Liouville space like $| \Omega )$ does. Cumulants need not be introduced here, since upgrades are connected. This explains why the EWP does not effect the MPS calculations for macroscopic chains. This holds true at least as long as an area law is holding. We hope that this topical review will help to give wavefunction based calculations for macroscopic systems a solid basis and to stimulate further work.

\section*{\Large Acknowledgements}
I would like to thank Younes Javanmard, Ingo Peschel, Frank Pollmann and Hermann Stoll for helpful discussions.

\end{document}